 \documentstyle[preprint,eqsecnum,aps]{revtex}

\begin{document}
\draft
\preprint{HEP/123-qed}
\title{Calculations on the Size Effects of Raman Intensities of\\ 
Silicon Quantum Dots}
\author{Wei Cheng$^{\dagger}$ and Shang-Fen Ren}
\address{Department of Physics, Illinois State University, Normal,
Illinois 61790-4560}

\date{\today}
\maketitle
\begin{abstract}
Raman intensities of Si quantum dots (QDs) with up to 
11,489 atoms (about 7.6 nm in diameter) for different scattering 
configurations are calculated. First, phonon modes in these QDs, 
including all vibration frequencies and vibration amplitudes,  
are calculated directly from the lattice dynamic matrix by using a 
microscopic valence force field model combined with the group theory. 
Then the Raman intensities of these quantum dots are 
calculated by using a bond-polarizability approximation. The size effects 
of the Raman intensity in these QDs are discussed in detail
based on these calculations. The calculations are compared with 
the available experimental observation. We are expecting 
that our calculations 
can further stimulate more experimental measurements.

\end{abstract}

\section{Introduction}
 
Semiconductor quantum dots (QDs) have attracted much
research attention in recent years because of their importance in
the fundamental understanding of physics and potential
applications in electronic devices, information processing, 
and non-linear optics. 
The electronic properties of QDs have been intensively studied in 
recent years, both theoretically and experimentally, and a clear 
understanding of much of the basic physics of the quantum confinement 
effects of electrons in QDs has been achieved\cite{Yoffe}. On the other
hand, the vibration properties of QDs, i.e., the confinement of
phonon modes in QDs, are less understood.

So far, most of the theoretical understanding of phonon modes in 
QDs are based on the continuum dielectric models. 
The analytic expression of the eigenfunctions of
LO phonons and surface optical phonons of small spherical
\cite{Duval,Klein,Frohlich,Fuchs,Ruppin,Giner,Roca,Cardona95,Cardona98} 
and cylindrical\cite{Li} QDs are derived and the electron-phonon
interactions are calculated. The extended 
continuum dielectric model\cite{Roca,Cardona95,Cardona98} coupling the
mechanical vibrational amplitudes and the electrostatic potential has
made major improvements over classical dielectric models in the study
of phonon modes in QDs. However, one of the basic assumptions of all
dielectric models is that the material is homogeneous and isotropic, 
that is
only valid in the long wavelength limit. When the size of QDs is
small, in the range of a few nm, the continuum dielectric models are
intrinsically limited. 

Many optical, transport, and thermal properties of quantum dots are 
related to phonon behavior in QDs. The theoretical treatment of these 
properties requires a reliable description of phonon modes and 
electron-phonon interaction potential in QDs. One of the major 
difficulties of the microscopic modeling of phonon modes in
QDs is its computational intensity. For example, for a GaAs QD of
size of about 8.0 nm, there are 11,855 atoms in it. Considering the three
dimensional motion of each atom, the dynamic matrix is in the order of
35,565. This is an intimidating task even with the most advanced
computers. In recent years, we have developed a microscopic valence 
force filed model (VFFM)\cite{QD1,QD2,QD3,QD4} to study phonon modes in 
QDs by employing the projection operators of the irreducible 
representations of the group theory to reduce the computational intensity. 
By employing the group theory, for example, the above matrix of size of 
35,565 can be reduced to five matrices in five different representations 
of A$_1$, A$_2$, E, T$_1$, and T$_2$, with the sizes of 1592, 
1368, 2960, 
4335, and 4560 respectively. Therefore, the original problem is reduced 
to a problem that can be easily handled by most reasonable computers. 
This allows not only the investigation of phonon modes in QDs with 
a much larger size, but also the investigation of phonon modes
in QDs with different symmetries. These investigations lead to many
interesting physics that otherwise can not be 
revealed\cite{QD1,QD2,QD3,QD4}. With this model, we have studied the size
effects of phonon modes in semiconductor quantum dots, including QDs 
of one material, such as GaAs or InAs, as well as QDs with a core of one 
material embedded in a shell of another material, such as GaAs cores 
embedded in AlAs shells. To further develop our theoretical model 
in investigations of properties of QDs, in this 
article, we have studied the size effects of Raman intensity in 
semiconductor QDs with the same model. 

It is well-known that the measurement of Raman spectra of a 
crystals is one of the most important methods for obtaining 
information about their lattice vibrations\cite{Loudon1,Loudon2}. 
Raman spectroscopy has been used to investigate the 
geometry, the nature, and the structures of QDs 
\cite{Duval,Klein,Yu2000,Freire,Hwang,Balandin,Armelles}. 
So far, most of the theoretical calculations on Raman scattering are 
based on phenomenological model \cite{Duval,Klein}. Recently there are 
calculations based on microscopic models \cite{JZ1,JZ2}, but because of 
the limitation on the computational intensity, the size range of the 
QDs that can be handled is limited.   

In this article, we have calculated 
the Raman intensities of nanoscale Si QDs by using the results of the VFFM 
together with a bond polarizability approximation 
(BPA) \cite{Bell,BFZ,JZ1,JZ2,Guha,Saito}. The 
calculated results are then compared  
with the available experimental data. We also hope that our 
calculations can further stimulate more experimental measurements on 
Raman intensities of semiconductor QDs.

The article is organized as the following. In Sections II, 
we describe the theoretical models of VFFM and BPA; in Section 
III, we show 
our calculated results of Raman intensities and have some discussions; and   
Section IV is a summary.  

\section{Theoretical Approaches}

\subsection{VFFM for Phonon modes in QDs}

In general, the theoretical model, VFFM, that we used to investigate 
phonon modes in QDs can be used to study phonon modes in groups IV, III-V, 
and II-VI semiconductors.  In this model, the change
of the total energy due to the lattice vibration is considered as
two parts, the change of the energy due to the short-range
interactions and the change of the energy due to the long range
Coulomb interaction:

\begin{equation}
\Delta E = \Delta E_s + \Delta E_c
\end{equation}
where the short-range interaction describes the covalent bonding, and
the long range part approximates the Coulomb interactions of polar 
semiconductor compounds. For the short range part,
we employed a VFFM with only two parameters as the
following\cite{Harrison}:

\begin{equation}
\Delta E_s=\sum\limits_{i}\frac{1}{2}C_0 (\frac {\Delta
d_i}{d_i})^2 +\sum\limits_{j}\frac{1}{2} C_1 (\Delta \theta_j)^2
\end{equation}
where $C_0$ and $C_1$ are two parameters describing the energy
change due to the bond length change and the bond angle change
respectively. The summation runs over all the bond lengths and
bond angles. Because each of these two parameters has a simple
and clear physical meaning, this model allows us to treat the
interaction between atoms near and at the surface appropriately. 
It can be further used to treat the effects of surface relaxations and 
reconstructions on the vibrations if necessary. The parameters $C_0$, 
$C_1$ used in our 
calculations for Si are 49.1 and 1.07 eV respectively \cite{Harrison}.
Since silicon is a homopolar semiconductor that all atoms are neutral, 
the long range Coulomb interaction is not necessary. 
                 
When considering the interaction between atoms, special attention
is paid to atoms near the surfaces of the QDs. More specifically,
for the short range interaction, when an atom is located near the
surface, interaction from its nearest neighboring atom is
considered only if that specific nearest atom is within the QD,
and interaction from its second neighboring atom is considered
only if that specific second neighbor atom is in the QD as well
as the nearest neighboring atom that makes the link between them.
The second point is important, because it makes sense with the
physics meanings of these two parameters, but it is easy to be
neglected.

As we discussed in above section, we have employed the projection
operators of the irreducible representations of the group theory
to reduce the computational intensity \cite{SYR,SYR2,SYR3,SYR4}
when calculating phonon modes directly from the dynamic matrices.
When the results of 
phonon modes are used to calculate the Raman intensity of QDs, the 
advantage of applying the group theory to calculate phonon modes in 
different symmetries is even more obvious. Because of the
symmetry dependence of Raman intensity, only phonon modes that are Raman 
active in that specific symmetry are necessary to be considered. This 
further reduces the amount of calculations required.

\subsection{Bond polarizability approximation}

With the vibrational normal modes obtained from the VFFM described,
we use the BPA \cite{Bell} to
calculate Raman scattering intensity of QDs. This BPA model was 
used in the prediction of Raman intensities of semiconductor
superlattices\cite{BFZ,JZ1,JZ2}, fullerences\cite{Guha}, and 
nanotubes \cite{Saito}. 

The BPA\cite{Bell} associates an axially symmetric polarizability tensor
with each bond as:

\begin{eqnarray}
\stackrel{\leftrightarrow }{P}(\vec{R}_{ij})=\alpha
(R_{ij})\stackrel{
\leftrightarrow }{I}+\gamma (R_{ij})[\hat{R}_{ij}\hat{R}_{ij}-\frac
{1}{3} \stackrel{\leftrightarrow }{I}]
\end{eqnarray}
where $\stackrel{\leftrightarrow }{I}$ is a unit matrix,
$\alpha(R_{ij})$ is the
mean polarizability, and $\gamma(R_{ij})$ describes the anisotropy of the
polarizability. Both $\alpha(R_{ij})$ and $\gamma(R_{ij})$ are functions
of the bond length $R_{ij}$ and not direction dependent.  Here
$\vec{R}_{ij}$ is the bond vector connecting atoms $i$ and $j$,
$R_{ij}$ is the length of $\vec{R}_{ij}$, and
$\hat{R}_{ij}=\vec{R}_{ij}/{R}_{ij}$. With phonon vibrations we have
$\vec{R}_{ij}=\vec{r}_{ij}+\vec{u}_{ij}=\vec{r}_{ij}
+\vec{u}_{j}-\vec{u}_{i}$,
where $\vec{r}_{ij}$ is the bond vector at equilibrium, $\vec{u}_{i}$ is
the displacement of atom $i$, and $\vec{u}_{ij}$ is the relative
displacement of atoms $j$ and $i$. For phonon vibrations, the condition 
that 
$\vec{u}_{ij}\ll\vec{r}_{ij}$ always applies. Therefore, the
polaribility tensor $\stackrel{\leftrightarrow}{P}(\vec{R}_{ij})$ 
can be expressed
in power of the displacements $\vec{u}_{ij}$ in a Talor's expansion. The
constant term of this expansion, i.e.,
$\stackrel{\leftrightarrow}{P}(\vec{r}_{ij})$,
can be ignored, since the total contribution of this term 
from all bonds 
to the Raman intensity 
vanishes. Keeping the first order of $\vec{u}_{ij}$ in the 
Taylor's expansion, the polarizability tensor 
$\stackrel{\leftrightarrow}{P}(\vec{R}_{ij})$ can
be simplified as the following:

\begin{eqnarray}
\stackrel{\leftrightarrow }{P}(\vec{R}_{ij})\approx (\vec{u}_{ij}\cdot
\hat{r}_{ij})[\alpha ^{\prime }(r_{ij})\stackrel{\leftrightarrow
}{I}+\gamma
^{\prime }(r_{ij})(\hat{r}_{ij}\hat{r}_{ij}-\frac 13\stackrel{
\leftrightarrow }{I})] \nonumber \\
+r_{ij}^{-1}\gamma(r_{ij})[\vec{u}_{ij}\hat{r}_{ij}+
\hat{r}_{ij}\vec{u}_{ij}-2(\vec{u}_{ij}\cdot
\hat{r}_{ij})\hat{r}_{ij}\hat{r}_{ij}]
\end{eqnarray}
where $\alpha^{\prime}(r_{ij})$ and $\gamma^{\prime }(r_{ij})$ are
derivatives of $\alpha(\vec{R}_{ij})$ and $\gamma(\vec{R}_{ij})$ 
with respect to $\vec{u}_{ij}$ evaluated at the equivalent bond-vector 
$\vec{r}_{ij}$ respectively, and $\hat{r}_{ij}=\vec{r}_{ij}/{r}_{ij}$.

In the zincblende structure, there are four different bond 
orientations, so we
only need to calculate $\stackrel{\leftrightarrow }{P}(\vec{R}_{ij})$
for these four different types of bonds. We can choose the following 
four directions as the bond orientations:

\begin{eqnarray}
\hat{r}_{01}=\frac 1{\sqrt{3}}(1,-1,1) \nonumber \\
\hat{r}_{02}=\frac 1{\sqrt{3}}(-1,1,1) \nonumber \\
\hat{r}_{03}=\frac 1{\sqrt{3}}(1,1,-1) \nonumber \\
\hat{r}_{04}=\frac 1{\sqrt{3}}(-1,-1,-1),
\end{eqnarray}

and the four polarizations associated with the four bonds are

\begin{eqnarray}
\stackrel{\leftrightarrow }{P}(\vec{R}_{01}) &=&\frac
1{\sqrt{3}}
(u_{01,x}-u_{01,y}+u_{01,z})[\alpha ^{\prime }\left(
\begin{array}{lll}
1 & 0 & 0 \\
0 & 1 & 0 \\
0 & 0 & 1
\end{array}
\right)+\frac 13\gamma ^{\prime }\left(
\begin{array}{lll}
0 & -1 & 1 \\
-1 & 0 & -1 \\
1 & -1 & 0
\end{array}
\right) ] \nonumber \\
& &+\frac \gamma {3\sqrt{3}}\left(
\begin{array}{lll}
4u_{01,x}+2u_{01,y}-2u_{01,z} & -u_{01,x}+u_{01,y}+2u_{01,z}
&
u_{01,x}+2u_{01,y}+u_{01,z} \\
-u_{01,x}+u_{01,y}+2u_{01,z} & -2u_{01,x}-4u_{01,y}-2u_{01,z}
&
2u_{01,x}+u_{01,y}-u_{01,z} \\
u_{01,x}+2u_{01,y}+u_{01,z} & 2u_{01,x}+u_{01,y}-u_{01,z}&
-2u_{01,x}+2u_{01,y}+4u_{01,z}
\end{array}
\right)
\end{eqnarray}

\begin{eqnarray}
\stackrel{\leftrightarrow }{P}(\vec{R}_{02}) &=&\frac
1{\sqrt{3}}
(-u_{02,x}+u_{02,y}+u_{02,z})[\alpha ^{\prime }\left(
\begin{array}{lll}
1 & 0 & 0 \\
0 & 1 & 0 \\
0 & 0 & 1
\end{array}
\right) +\frac 13\gamma ^{\prime }\left(
\begin{array}{lll}
0 & -1 & -1 \\
-1 & 0 & 1 \\
-1 & 1 & 0
\end{array}
\right) ] \nonumber \\
& &+\frac \gamma {3\sqrt{3}}\left(
\begin{array}{lll}
-4u_{02,x}-2u_{02,y}-2u_{02,z} & u_{02,x}-u_{02,y}+2u_{02,z}
&
u_{02,x}+2u_{02,y}-u_{02,z} \\
u_{02,x}-u_{02,y}+2u_{02,z} & 2u_{02,x}+4u_{02,y}-2u_{02,z}
& 2u_{02,x}+u_{02,y}+u_{02,z} \\
u_{02,x}+2u_{02,y}-u_{02,z} & 2u_{02,x}+u_{02,y}+u_{02,z}
& 2u_{02,x}-2u_{02,y}+4u_{02,z}
\end{array}
\right)
\end{eqnarray}

\begin{eqnarray}
\stackrel{\leftrightarrow }{P}(\vec{R}_{03}) & =&\frac
1{\sqrt{3}}
(u_{03,x}+u_{03,y}-u_{03,z})[\alpha ^{\prime }\left(
\begin{array}{lll}
1 & 0 & 0 \\
0 & 1 & 0 \\
0 & 0 & 1
\end{array}
\right) +\frac 13\gamma ^{\prime }\left(
\begin{array}{lll}
0 & 1 & -1 \\
1 & 0 & -1 \\
-1 & -1 & 0
\end{array}
\right) ] \nonumber \\
& & +\frac \gamma {3\sqrt{3}}\left(
\begin{array}{lll}
4u_{03,x}-2u_{03,y}+2u_{03,z} & u_{03,x}+u_{03,y}+2u_{03,z}
& -u_{03,x}+2u_{03,y}+u_{03,z} \\
u_{03,x}+u_{03,y}+2u_{03,z} & -2u_{03,x}+4u_{03,y}+2u_{03,z}
& 2u_{03,x}-u_{03,y}+u_{03,z} \\
-u_{03,x}+2u_{03,y}+u_{03,z} & 2u_{03,x}-u_{03,y}+u_{03,z}
& -2u_{03,x}-2u_{03,y}-4u_{03,z}
\end{array}
\right)
\end{eqnarray}

\begin{eqnarray}
\stackrel{\leftrightarrow }{P}(\vec{R}_{04}) & = & \frac
1{\sqrt{3}}
(-u_{04,x}-u_{04,y}-u_{04,z})[\alpha ^{\prime }\left(
\begin{array}{lll}
1 & 0 & 0 \\
0 & 1 & 0 \\
0 & 0 & 1
\end{array}
\right) +\frac 13\gamma ^{\prime }\left(
\begin{array}{lll}
0 & 1 & 1 \\
1 & 0 & 1 \\
1 & 1 & 0
\end{array}
\right) ] \nonumber \\
& &+\frac \gamma {3\sqrt{3}}\left(
\begin{array}{lll}
-4u_{04,x}+2u_{04,y}+2u_{04,z} & -u_{04,x}-u_{04,y}+2u_{04,z}
& -u_{04,x}+2u_{04,y}-u_{04,z} \\
-u_{04,x}-u_{04,y}+2u_{04,z} & 2u_{04,x}-4u_{04,y}+2u_{04,z}
& 2u_{04,x}-u_{04,y}-u_{04,z} \\
-u_{04,x}+2u_{04,y}-u_{04,z} & 2u_{04,x}-u_{04,y}-u_{04,z}
& 2u_{04,x}+2u_{04,y}-4u_{04,z}
\end{array}
\right).
\end{eqnarray}

\vbox{}
When we calculate the polarizability tensor for the QDs,
we sum the polarizability tensor associated with each bond in the QDs,
and get a scattering tensor 

\begin{eqnarray}
\stackrel{\leftrightarrow}{P}={\sum_{i<j} }
\stackrel{\leftrightarrow }{P}(\vec{R}_{ij}).
\end{eqnarray}

From the symmetry property, we know that the Raman tensor in cubic
crystals takes the following forms\cite{Yu}:
                                            
\begin{eqnarray}
\stackrel{\leftrightarrow }{P}_{A1}=\left(
\begin{array}{ccc}
a & 0 & 0 \\
0 & a & 0 \\
0 & 0 & a
\end{array}
\right)
\end{eqnarray}

\begin{eqnarray}
\stackrel{\leftrightarrow }{P}_E=\left(
\begin{array}{ccc}
b & 0 & 0 \\
0 & b & 0 \\
0 & 0 & -2b
\end{array}
\right) ;\left(
\begin{array}{ccc}
\sqrt{3}b & 0 & 0 \\
0 & -\sqrt{3}b & 0 \\
0 & 0 & 0
\end{array}
\right)
\end{eqnarray}

\begin{eqnarray}
\stackrel{\leftrightarrow }{P}_{T2}=\left(
\begin{array}{ccc}
0 & 0 & 0 \\
0 & 0 & d \\
0 & d & 0
\end{array}
\right) ;\left(
\begin{array}{ccc}
0 & 0 & d \\
0 & 0 & 0 \\
d & 0 & 0
\end{array}
\right) ;\left(
\begin{array}{ccc}
0 & d & 0 \\
d & 0 & 0 \\
0 & 0 & 0
\end{array}
\right).
\end{eqnarray}

In our calculations we found that for semiconductor QDs with 
zincblende structure, after we applied the projection operators, 
$\stackrel{\leftrightarrow }{P}$ is a constant times a 
unit matrix for an A$_1$ mode, a traceless diagonal matrix with two
equal matrix elements for an E mode, and a
traceless matrix with only non-diagonal matrix elements for a T$_2$ mode.
The Raman tensors $\stackrel{\leftrightarrow }{P}$ for the QDs can 
be expressed as the following forms:

\begin{eqnarray}
\stackrel{\leftrightarrow }{P}_{A1}=\left(
\begin{array}{ccc}
A_l & 0 & 0 \\
0 & A_l & 0 \\
0 & 0 & A_l
\end{array}
\right) 
\end{eqnarray}

\begin{eqnarray}
\stackrel{\leftrightarrow }{P}_{E}=\left(
\begin{array}{ccc}
2E_l & 0 & 0 \\
0 & -E_l & 0 \\
0 & 0 & -E_l
\end{array}
\right) 
\end{eqnarray}

\begin{eqnarray}
\stackrel{\leftrightarrow }{P}_{T2}=\left(
\begin{array}{ccc}
0 & 0 & 0 \\
0 & 0 & T_l \\
0 & T_l & 0
\end{array}
\right)
\end{eqnarray}
where 
\begin{eqnarray}
A_l^2=[(P_{11}^l+P_{22}^l+P_{33}^l)/3]^2, 
\end{eqnarray}
\begin{eqnarray}
E_l^2=\frac{1}{18}[(P_{11}^l-P_{22}^l)^2+(P_{22}^l-P_{33}^l)^2+
(P_{33}^l-P_{11}^l)^2],
\end{eqnarray}
and 
\begin{eqnarray}
T_l^2=[(P_{12}^l)^2+(P_{23}^l)^2+(P_{31}^l)^2], 
\end{eqnarray}
where $P_{ij}^l$ is the element of the change of the polarizability 
resulting from vibration of the QDs in mode $l$ as defined in (2.10).
   
The $A_l^2$, $E_l^2$, and $T_l^2$ are invariants of the Raman 
tensor and $l$ is an index of the modes. This is 
consistent with the Raman tensor in cubic crystals of 
above (2.11 - 2.13). Here $A_l^2$ is related to the intensity of an
A$_1$ mode in polarized Raman geometry, and $E_l^2$ and $T_l^2$ 
are related to intensities of E and T$_2$ modes in depolarized Raman 
geometry respectively. For E or T$_2$ 
modes that are two- or three-fold degenerated, the contributions 
of the degeneracies are considered in the total scattering 
intensity\cite{Loudon1,Loudon2}. 

For unpolarized incident light with frequency $\Omega$ scattered 
perpendicular to the
direction of propagation, the intensities of Raman scattered 
components with frequencies $ \Omega \pm \omega _l$ are the 
following\cite{Bell}:

\begin{eqnarray}
I_{||}=\frac{(\bar{n}_l+1/2\pm 1/2)}{2\omega 
_l}g(\omega_l)[7G_l^2+45A_l^2]
\end{eqnarray}

\begin{eqnarray}
I_{\perp }=\frac{(\bar{n}_l+1/2\pm 1/2)}{2\omega_l}g(\omega_l)[6G_l^2]
\end{eqnarray}
where $\bar{n}_l$ is the average occupation number of phonon mode 
$l$, and $I_{||}$ and $I_{\perp }$ are intensities of scattered light
with polarization parallel and perpendicular to the plane of 
scattering. In these equations,  $A_l^2$ is the same as defined 
in (2.17), and 

\begin{eqnarray}
G_l^2=9E_l^2+3T_l^2.
\end{eqnarray}

The Raman scattering matrix for cubic
crystals with arbitrary orientations and arbitrary
incident and scattering light wave vector are studied  
by using Stokes vectors \cite{Chan}. The Stokes vectors are 
defined as the following: 
 
Let the incident elliptic polarization be described as

\begin{eqnarray}
\vec{E}=(a+ib)\vec{E}_A+(c+id)\vec{E}_N,
\end{eqnarray}
where $\vec{E}_A$ and $\vec{E}_N$ are 
polarization vectors in and normal to the scattering plane. These 
two directions are labeled as $\vec{A}$ and $\vec{N}$ 
respectively. Its Stokes vector is defined as

\begin{eqnarray}
\left(
\begin{array}{c}
a^2+b^2+c^2+d^2 \\
a^2+b^2-c^2-d^2 \\
2(ac+bd) \\
2(ad-bc)
\end{array}
\right).
\end{eqnarray}

For example, for a left handed circularly polarized light
viewed along the direction of propagation of 
the light, its polarization vector is $\vec{E}=\frac 
i{\sqrt{2}}\vec{E}_A+\frac 
1{\sqrt{2}}\vec{E}_N$, and its equivalent Stokes vector 
is (1, 0, 0, -1). From this, 
the Stokes parameters of the scattering light are derived choosing four
different combinations of values of $a$, $b$, $c$, and $d$. This 
gives the complete scattering matrix $a_{ij}$. Assume that QDs orient
randomly, all the possible incident directions should be averaged. 
Then the scattering matrix for A$_1$ modes is\cite{Chan}

\begin{eqnarray}
a_{ij}=
\left(
\begin{array}{cccc}
1+\cos ^2\Theta & -1+\cos ^2\Theta & 0 & 0 \\
-1+\cos ^2\Theta & 1+\cos ^2\Theta & 0 & 0 \\
0 & 0 & 2\cos \Theta & 0 \\
0 & 0 & 0 & 2\cos \Theta
\end{array}
\right), 
\end{eqnarray}
and for both E and T$_2$ modes it is

\begin{eqnarray}
a_{ij}=
\left(
\begin{array}{cccc}
13+\cos ^2\Theta & -1+\cos ^2\Theta & 0 & 0 \\
-1+\cos ^2\Theta & 1+\cos ^2\Theta & 0 & 0 \\
0 & 0 & 2\cos \Theta & 0 \\
0 & 0 & 0 & -10\cos \Theta
\end{array}
\right), 
\end{eqnarray}
where  $\Theta $ is the scattering angle.
                                                                           
In our calculations, we have calculated $A_l^2$, 9$E_l^2$, and 3$T_l^2$ 
according to the right-angle scattering equations (2.20-2.22). For 
different incident light and scattering configurations, the Raman 
intensities are the linear combination of these three according to their 
Stokes vectors.  

\section{Results and Discussion}

In this article we report calculated reduced Raman scattering 
intensity, $\omega _lI/(\bar{n}_l+\frac 12\pm \frac 12)$, 
for Si QDs with diameters from 15 to 76  \AA. Since 
$\alpha^{\prime }$, 
$r_{ij}^{-1}\gamma $, and $\gamma ^{\prime }$ are not accurately known 
a priori, these parameters used in our calculations satisfy\cite{Bell}

\begin{eqnarray} 
r_{ij}^{-1}\gamma (r_{ij})=\frac 38\gamma^{\prime 
}(r_{ij})=\frac 38\alpha^{\prime}(r_{ij})
\end{eqnarray}

We noticed that in the early calculations of amorphous Si 
\cite{Bell} $\alpha^{\prime}\simeq 0$ was assigned
because the observed intensity profiles $I_{||}$ and $I_{\perp }$ 
of Si have the same shape. If $\alpha^{\prime}=0$, there 
would be no contribution to Raman intensity from A$_1$ modes. 
However, when the size of QDs is small,  
the contribution from A$_1$ modes might be important to 
consider, so we have assigned $\alpha^{\prime}$ a number as 
above. The choice of this number is not critical, since for 
different choices of $\alpha^{\prime}$, the shape of 
Raman intensity from A$_1$ contribution is the same, and only its 
relevant strength to E and T$_2$ modes is different.  
 
We have calculated Raman intensities of A$_1$, E, and T$_2$ modes 
for Si QDs with approximate sizes of 15, 20, 25, 30, 35, 40, 
50, 60, 70, and 76 \AA { }
respectively, and the results are shown in Figs. 1-3. 
Here the Raman intensity $I$ is calculated by the Lorentz 
broadening 
\begin{eqnarray}
I=\sum\limits_l\frac{I_l{\Gamma/\pi}}{(\Gamma)^2+(\omega -\omega _l)^2},
\end{eqnarray}
where $I_l$ and $\omega _l$ are the scattering intensity and
eigenfrequency of mode $l$ respectively, and $\Gamma$ is the half 
Lorentz width, which is taken as $2/\pi=0.64$ $cm^{-1}$ in our 
calculations.   
We have also listed the related important data in Table 1
that are numerically more clear. The 
data listed in Table 1 in order are the diameters of the QDs 
calculated ($d$), the number of atoms in the QDs ($N$), the 
intensity of the 
highest peak of A$_1$ modes ($I_{A1}$), the intensity of the 
highest peak of E modes ($I_{E}$), the intensity of the highest peak of 
T$_2$ modes in the low frequency range ($I_{T2}(l)$) and in the 
high frequency range ($I_{T2}(h)$), the frequency of the first peak of
A$_1$ modes ($\omega_{A1}$), the frequency of the first peak of
E modes ($\omega_{E}$), the frequency of the first peak of
T$_2$ modes in the low frequency range ($\omega_{T2}(l)$), and the 
frequency of the highest peak of T$_2$ modes in the high 
frequency range ($\omega_{T2}(h)$).
All the intensities listed above are Raman intensity per atom for easier 
comparison. We will discuss the important 
features of these results in detail next.

\subsection{Size effects of highest frequencies}

From Figs. 1-3 we see that in general, the major peaks in the high 
frequency range always have a T$_2$ symmetry. The highest peaks correspond 
to T$_2$ phonon modes with the highest frequencies. When 
the size of the QDs increases, this frequency approaches the frequency 
of the optical phonon frequency of bulk Si. Theoretically speaking, 
when the size of QDs approaches infinite, the total Raman spectrum of 
QDs approaches the Raman spectrum of bulk Si, and this will be the 
only peak left.  

In our previous calculations of phonon modes in QDs\cite{QD1,QD2,QD3,QD4}, 
we have 
discussed the size dependance of phonon modes with different symmetries. We 
have learned that phonon modes with different symmetries have different 
size dependance, and A$_1$ modes usually have the strongest size effect. 
For quantum dots of zincblend semiconductors, such as GaAs, 
when the size of QDs is big, the mode with the highest frequency has
A$_1$ symmetry. However, as the dot size decreases, there 
is a crossover of the symmetries, then the mode with the highest frequency 
has T$_2$ symmetry.  For Si quantum dots studied in the 
present paper, the mode with the highest frequency is always of A$_1$ 
symmetry. However, the Raman intensity of A$_1$ modes decreases  
when the dot size increases, so the strongest high frequency mode is 
always of T$_2$ symmetry. 

When the size of QDs decreases, the frequency of this T$_2$ peak 
decreases. 
To show this more clearly, we enlarged the high frequency range of 
Fig. 3 and plotted it as
Fig. 4. From the data listed in Table 1, we know that when the 
diameters of Si QDs decreases from 75.79 \AA { }   
to 14.11 \AA, the frequency of the 
highest Raman peak shifts from 518.3 $cm^{-1}$ to 506.4 $cm^{-1}$
(the Raman peak of Bulk Si is at 518.9 $cm^{-1}$ in our model). 
The systematic redshift of the 
longitudinal (LO) phonon peaks due to spacially confined 
phonon modes in nanocrystals in the 
size range of a few nm have been observed 
\cite{Freire,Hwang,Balandin}, and 
recently it has been observed by resonant Raman scattering in three 
samples of Ge 
nanocrystals in the size range of 4-10 nm \cite{Yu2000}.     

One more thing we notice from Fig. 3 and Fig. 4 for the 
high frequency peaks of T$_2$ modes is that not only the highest 
intensity peak red shifts as dot size decreases, 
but also weaker peaks appear at the same time. 

Experimentally it may be difficult to resolve all the weaker peaks because 
of broadening resulting from fluctuation in dot sizes. As a result one may 
observe an asymmetric broadening of the Raman peak corresponding to the 
optical phonon as the dot size is reduced.
This is indeed found in Raman intensities of Ge QDs \cite{Yucom1}. 
One may attempt to inteprete this as an indication that the quality
of the dots may be poorer leading to larger inhomogeneous broadening
as dot size gets smaller. However, from our calculations on Raman 
intensities of QDs, one can
notice that the red shift of the strongest T$_2$
Raman peak is smaller than the frequency spread of the weaker peaks
which appear. In other words the broadening of the Raman peak is
larger than the red shift as the dot size decreases. This indicates that 
the observed broadening in Raman measurements is not only due to the red 
shift of the peak alone, but there is also a contribution to this 
broadening from quantum size effect.     

\subsection{Size effects of lowest frequencies}

From Figs. 1-3, we see that the frequency of the first peak in the low 
frequency range for all three different symmetries (A$_1$, E, and T$_2$) 
increases as the size of the QDs decreases. From the data listed in Table 
1, we see that when the size of Si QDs decreases from 75.79 \AA 
{ } to 14.11 \AA, the frequency of the first Raman peak of A$_1$ modes 
shifts from 28.6 $cm^{-1}$ to 129.8 $cm^{-1}$, the first Raman peak of 
E modes shifts from 8.6 $cm^{-1}$ to 46.7 $cm^{-1}$, and the first Raman 
peak with T$_2$ symmetry shifts from 12.4 $cm^{-1}$ to 58.8 $cm^{-1}$.
The size effects of lowest frequencies of phonon modes in QDs 
have been discussed in detail 
in our previous studies\cite{QD1,QD2,QD3,QD4}, and again this is
shown in the calculated Raman spectra. Furthermore, we 
notice that even though the
frequencies of the lowest frequency peak increase in all these three  
figures, they increase at a different rate. The lowest frequency of the 
A$_1$ peak increases much faster than that of the other two. 
To show this more clearly, we plot the 
lowest frequency peaks versus the sizes of QDs in Fig. 5.
It is obvious that the lowest frequency of the A$_1$ peak 
increases much faster than 
that of the other two, because that the A$_1$ modes have the strongest 
quantum confinement effects\cite{QD1,QD2}. 

Another feature of Fig. 5 is that the lowest frequencies of the Raman 
peaks 
in the acoustic range are roughly proportional to the inverse of the QDs 
diameters, which was first observed by Duval and his co-workers 
\cite{Duval}. This was recently observed in Si nanocrystals
\cite{Fujii}, and it was noticed that the depolarized Raman 
spectra appear at much lower frequencies than the polarized ones.
Not only our results agree with the experimental observations, we 
also learned that this is actually due to the symmetry dependence of 
the confinement effect of phonon modes, i.e., the A$_1$ modes have 
the strongest confinement effects.       
 
\subsection{Folding of the acoustic phonons}

In Fig. 1, we see that at the low frequency range of the A$_1$ modes, 
the Raman spectra is dominated by a series of nearly evenly spaced 
peaks in the acoustical phonon range. As the size decreases, the spacing 
increases. This can be understood from the folding of acoustic 
phonons. Since the A$_1$ modes vibrate in the radical direction, 
when the radius of the QDs increases approximately one lattice constant, 
there will be one more folding due to the confinement of the QDs along the 
radial direction. This should be observable by Raman 
scattering, and we are expecting such observations. 
   
\subsection{Size effects on strength of Raman peaks}

Since Raman intensities plotted in Figs. 1-3 are in arbitrary unit, the 
size effects on strength of Raman peaks are not shown clearly in these 
figures.  To show it more clearly, we 
have plotted the Raman intensity per atom versus diameters of QDs 
for the low frequency A$_1$ peaks, the low frequency E peaks, the low 
frequency T$_2$ peaks, and the high frequency T$_2$ peaks in Fig. 6. 
For A$_1$ Raman spectra there are several high peaks, and we choose
the strength of the highest peak (when the size is less than 30 \AA, 
this peak is not the peak with the lowest frequency). In Fig. 6 we see 
that the 
strength of low frequency peaks (A$_1$, E, and T$_2$) decreases 
fast as the size of QDs increases, and the strength of high frequency 
peaks (T$_2$) remains a constant in QDs of all sizes. This indicates 
that even though in bulk material only one major 
peak can be measured, when the size of QDs decreases, other 
peaks in the low frequency range will appear.  Of these low frequency 
Raman peaks, the most noticable ones are probably the evenly 
spaced A$_1$ peaks in the polarized spectra. Such evenly spaced A$_1$ 
peaks should be observable.

We want to emphasize that in Fig. 6, the Raman intensity shown is 
from one calculated highest Raman peak, either in the low frequency 
range or in the high frequency range.  Experimentally the presence of 
multiple Raman 
peaks as shown in Figs. 1-4 may not be resolvable due to size
fluctuation, and in stead a broadened peak is observed.
Typically in such situation all the Raman intensity
should be intergrated together to obtain the total
strength of the Raman peak. To compare with the experimental results, 
we summed the calculated 
Raman intensities (without broadening) for all A$_1$, E, and T$_2$ modes 
respactively and show them for QDs with different sizes in Fig. 7. We can 
see that 
all the Raman intensities will increase when the size of QDs get smaller. 
When the QDs get larger, the Raman intensity of T$_2$ mode will approach 
to the Raman intensity of bulk crystal and others will approach to zero.  
This is in qualitative agreement with what was observed
in Ge QDs \cite{Yucom1}. 

%One more comment we want to make is about the intensity of high 
%frequency T$_2$ peak. Even though our Fig. 6  
%shows that the highest T$_2$ peak intensity is almost a constant and 
%independent of dot size, this is only the contribution from one 
%peak. 
%
%Experimentally the presence of several Raman peaks as shown in Figs. 
%3 and 
%4 as the dot size becomes smaller may not be resolvable due to size 
%fluctuation and in stead a broadened peak is observed. 
%Typically in such situation all the Raman intensity 
%should be intergrated together to obtain the total
%strength of the Raman peak. From Fig. 3 and Fig. 4 we see that 
%there are more peaks of 
%high frequency T$_2$ modes when the size of quantum QDs is small, 
%therefore if we integrate the
%intensity of all T$_2$ peaks we will see that the total strength
%actually also increases as the dot size decreases just like the
%other low frequency peaks. The only difference is that the highest 
%T$_2$ peak will appear for bulk crystal but not those low frequency 
%peaks.  This is in qualitative agreement with what was observed 
%in Ge QDs \cite{Yucom1}.      

\subsection{Size effects on mode mixing}

One more thing we want to comment on from Figs. 1-3 and Fig. 6 is that for 
large size QDs, the major peak of the Raman spectra is derived from the 
T$_2$ high frequency mode. This peak approaches the optical phonon peak 
in the 
bulk Raman spectra when the size of QDs is large. When the size of QDs 
decreases, more and 
stronger peaks at the lower frequency range show up, which are 
derived from the A$_1$ modes in the polarized Raman spectra and 
E and T$_2$ modes from the depolarized Raman spectra. As can be 
seen from Figs. 1-3 and Fig. 6, the intensities of the
A$_1$, E, and T$_2$ modes are of nearly the same
magnitude for small dots, which indicates the 
mode mixing due to the quantum confinement of phonon modes
in small QDs.   
 
\section{Summary}

In summary, we have calculated the Raman intensities of Si 
QDs with up to 11,489 atoms (about 7.6 nm in 
diameter). The phonon modes are calculated directly from the 
lattice dynamic matrix of a microscopic 
VFFM by employing the projection operators of the irreducible 
representations. Based on the results of phonon modes, the Raman 
intensities are calculated by using a BPA. The size effects 
of the Raman intensity in QDs are discussed in detail based 
on these calculations. Our calculated results agree with the 
existing experimental observations, and we are expecting that our 
calculations will stimulate 
more experimental measurements of Raman intensities of QDs.
               
\acknowledgements

This research is supported by the National Science Foundation
(DMR9803005 and INT0001313). We thank Profs. Shang-Yuan Ren
and Bang-Fen Zhu  for helpful discussions. We want to thank Prof. 
Peter Yu for helpful comments that provide us the experimental details 
on Ge QDs and point us further to the direction of comapring our 
calculated results with the experimental observations.     
W. Cheng is grateful to Illinois State 
University for hosting his visit.

\mbox{}\\
{}$^{\dagger}$ On leave from the Institute of Low Energy Nuclear Physics, 
Beijing Normal University, Beijing, 100875, P. R. China.

\newpage

\begin{figure}
\caption
{Reduced Raman intensities of A$_1$ modes
for Si QDs with approximate diameters in \AA {} indicated.}
\label{fig1}
\end{figure}
             
\begin{figure}
\caption
{Reduced Raman intensities of E modes
for Si QDs with approximate diameters in \AA {} indicated.}
\label{fig2}
\end{figure}

\begin{figure}
\caption
{Reduced Raman intensities of T$_2$ modes
for Si QDs with approximate diameters in \AA {} indicated.}
\label{fig3}
\end{figure}

\begin{figure}
\caption
{Reduced Raman intensities of T$_2$ modes
enlarged at the high frequency range
for Si QDs with approximate diameters in \AA {} indicated.}
\label{fig4}
\end{figure}

\begin{figure}
\caption
{Frequency of the lowest Raman peak of A$_1$, E, and T$_2$ 
modes versus size of the dots for Si QDs.}
\label{fig5}
\end{figure}

\begin{figure}
\caption
{Raman intensity per atom of the highest low-frequency peaks of A$_1$, E, 
and T$_2$ modes  and the highest high frequency peaks of T$_2$ modes 
versus size of the dots for Si QDs.}
\label{fig6}
\end{figure}

\begin{figure}
\caption
{Integrated Raman intensity per atom of 
A$_1$, E,
and T$_2$ modes versus size of the dots for Si QDs.}
\label{fig7}
\end{figure}

\newpage
\begin{table}
\caption{Raman intensities of A$_1$, E, and T$_2$ modes
for Si QDs with approximate sizes of 15, 20, 25, 30, 35, 40,
50, 60, 70, and 76 \AA. The data listed in order are the 
diameters of the QDs ($d$), the number of atoms in the QDs ($N$), 
the intensity of the highest peak of A$_1$ modes ($I_{A1}$), 
the intensity of the highest peak of E modes ($I_{E}$), 
the intensity of the highest peak of T$_2$ modes in the low 
frequency range ($I_{T2}(l))$, 
the intensity of the highest peak of T$_2$ modes in the
high frequency range ($I_{T2}(h))$, 
the frequency of the first peak of A$_1$ mode ($\omega_{A1}$), 
the frequency of the first peak of E mode ($\omega_{E}$),
the frequency of the first peak of T$_2$ mode in the low frequency range 
($\omega_{T2}(l)$), and 
the frequency of the highest peak of T$_2$ mode in the high
frequency range ($\omega_{T2}(h)$).
All the intensities listed here are Raman intensity per 
atom.} \label{tabel1}

\vbox{}

\begin{tabular}{llllllllll}

$d$ & $N$ & $I_{A1}$ & $I_E$ & $I_{T2}(l)$ & $I_{T2}(h)$ & $\omega
_{A1}$ & $\omega _E$ & $\omega _{T2}(l)$ & $\omega _{T2}(h)$ \\ \hline
14.11& 87  & 7.65 & 9.70 &10.99 & 9.88 & 129.8 & 46.7 & 58.8 & 506.4 \\
19.39& 191 & 5.88 & 5.58 & 6.61 & 9.60 & 98.4 & 35.2 & 46.9 & 511.2 \\
24.74& 417 & 3.55 & 3.53 & 4.79 & 9.47 & 83.3 & 27.2 & 37.5 & 514.3 \\
29.75& 705 & 2.69 & 2.76 & 3.66 & 9.22 & 69.0 & 23.3 & 31.2 & 515.6 \\
34.67& 1099& 2.46 & 2.03 & 2.64 & 9.25 & 61.4 & 19.7 & 26.6 & 516.5 \\
39.91& 1707& 1.84 & 1.51 & 1.97 & 9.24 & 53.2 & 16.9 & 22.9 & 517.1 \\
50.00& 3265& 1.27 & 1.01 & 1.34 & 9.47 & 43.2 & 13.4 & 18.8 & 517.7 \\
59.99& 5707& 0.91 & 0.72 & 0.93 & 9.67 & 36.0 & 10.9 & 15.5 & 518.1 \\
69.97& 9041& 0.68 & 0.54 & 0.69 &10.04 & 30.9 &  9.2 & 13.4 & 518.3 \\
75.79& 11489&0.58 & 0.46 & 0.59 &10.26 & 28.6 &  8.6 & 12.4 & 518.3 \\

\end{tabular}
\end{table}


\begin{references}

\bibitem{Yoffe} A. D. Yoffe, Adv. Phys. 42, 173 (1993); A. D. Yoffe, 
Adv. Phys. 50, 1 (2001).

\bibitem{Duval} E. Duval, Phys. Rev. B 46, 5795 (1992).

\bibitem{Klein} M. C. Klein, F. 
Hache, D. Ricard and C. Flytzanis, Phys. Rev. B 42, 11123 (1990).

\bibitem{Frohlich} H. Frohlich, Oxford University Press, Oxford (1949).

\bibitem{Fuchs} R. Fuchs and K. L. Kliewer, Phys. Rev. 140, A 2076 
(1965).

\bibitem{Ruppin} R. Ruppin, and R. Englman, Rep. Prog. Phys. 33, 144
(1970).

\bibitem{Giner} C. Trallero-Giner, F. Garcia-Moliner, V. R. Velasco, and
M. Cardona, Phys. Rev. B 45, 11944 (1992).

\bibitem{Roca} E. Roca, C. Trallero-Giner, and M. Cardona, Phys. Rev. B 
49, 13704 (1994).

\bibitem{Cardona95} M. P. Chamberlain, C. Trallero-Giner, and M. 
Cardona,
Phys. Rev. B 51, 1680 (1995).

\bibitem{Cardona98} C. Trallero-Giner, A. Debernardi, M. Cardona, E.
Menendez-Proupin, and A. I. Ekimov, Phys. Rev. B 57, 4664 (1998).

\bibitem{Li} W. S. Li and C. Y. Chen, Physica B 229, 375 (1997).

\bibitem{QD1} S. F. Ren, Z. Q. Gu and D. Y. Lu. Solid 
State Comm. 113, 273 (2000).

\bibitem{QD2} S. F. Ren, D. Y. Lu, and G. Qin, Phys. Rev. B 63, 195315 
(2001).

\bibitem{QD3} G. Qin and S. F. Ren, J. Appl. Phys. 89 (11), 6037 (2001).

\bibitem{QD4} G. Qin and S. F. Ren, to be published on Soild State 
Comm. (2001).

\bibitem{Loudon1} R. Loudon, Proc. Roy. Soc. London, 275, 218 (1963).

\bibitem{Loudon2} R. Loudon, Adv. Phys. 13, 423 (1964); R. 
Loudon, Adv. Phys. 14, 621 (1965).

\bibitem{Yu2000} K. L. Teo, S. H. Kwok, P. Y. Yu, and S. Guha, 
Phys. Rev. B 62, 1584 (2000).

\bibitem{Freire} P. T. C. Freire, M. A. Araujo Silva, V. C. S. 
Reynoso, A. R. Vaz, and V. L. Lemos, Phys. 
Rev. B 55, 6743 (1997).

\bibitem{Hwang} Y. N. Hwang, S. Shin, H. L. Park, S. H. Park, 
U. Kim, H. S. Jeong, E. J. Shin, and D. Kim, Phys. Rev. B 54, 15120 
(1996).

\bibitem{Balandin} A. Balandin, K. L. Wang, N. Kouklin and S. 
Bandyopadhyay, Appl. Phys. Lett. 76, 137 (2000).

\bibitem{Armelles} G. Armelles, T. Utzmeier, P. A. Postigo, F. 
Briones, J. C. Ferrer, P. Peiro, and A. Cornet, J. Appl. Phys. 
81, 6339 (1997).

\bibitem{Bell} R. J. Bell, in Methods in Computational Physics,
edited by B. Alder, S. Fernbach, and M. Rotenberg 
(Academic, New York, 1976), Vol. 15, P.260.

\bibitem{BFZ} B. F. Zhu and K. A. Chao, Phys. Rev. B 36, 4906 (1987).

\bibitem{JZ1} J. Zi, H. Buscher, C. Falter, W. Ludwig, 
K. Zhang, and X. Xie, Appl. Phys. Lett. 69, 200 (1996).

\bibitem{JZ2} J. Zi, K. Zhang, and X. Xie, Phys. Rev. B 58, 6712 (1998).

\bibitem{Guha} S. Guha, J. Menendez, J. B. Page, and G. B. Adams,
Phys. Rev. B 53, 13106 (1996).

\bibitem{Saito} R. Saito, T. Takeya, and T. Kimura, G. Dresselhaus, 
and M. S. Dresselhaus, Phys. Rev. B 57, 4145 (1998).

\bibitem{Harrison} W. A. Harrison, Electronic Structure and the 
Properties of Soilds, Freeman, San Francisico, 1980.

\bibitem{Kunc} K. Kunc, M. Balkanski, and M. A. Nusimovici, 
Phys. Stat. Sol. (b) 72, 229 (1975); K. Kunc, Ann. Phys. (France) 
8, 319 (1973-1974).

\bibitem{SYR} S. Y. Ren, Phys. Rev. B 55, 4665 (1997).

\bibitem{SYR2} S. Y. Ren, Solid State Comm. 102, 479 
(1997).

\bibitem{SYR3} S. Y. Ren, Jpn. J. Appl. Phys. 36, 3941 
(1997).

\bibitem{SYR4} S. Y. Ren and S. F. Ren, J. Phys. Chem. 
Solid 59, 1327 (1998).

\bibitem{Yu}
P. Yu and M. Cardona, Fundamentals of Semiconductors, 
Physics and Materials Properties, Springer, Berlin, 1996.

\bibitem{Chan}  V. 
Chandrasekharan, Z. Phys. 175, 63 (1963).

\bibitem{Yucom1} P. Yu, private communication.

\bibitem{Fujii}M. Fujii, Y. Kanzawa, S. Hayashi, and K. Yamamoto,
Phys. Rev. B 54, R8373 (1996).  

%\bibitem{Kobayashi} T. Kobayashi, T. Endoh, H. Fukuda, and S. 
%Nomura, A. Sakai, and Y. Ueda, Appl. Phys. Lett. 71, 1195 
%(1997). 

%\bibitem{Paine} D. C. Paine, C. Caragianis, T. Y. Kim, 
%Y. Shigesato, and T. Ishahara, Appl. Phys. Lett. 62, 2842 (1993).

\end{references}
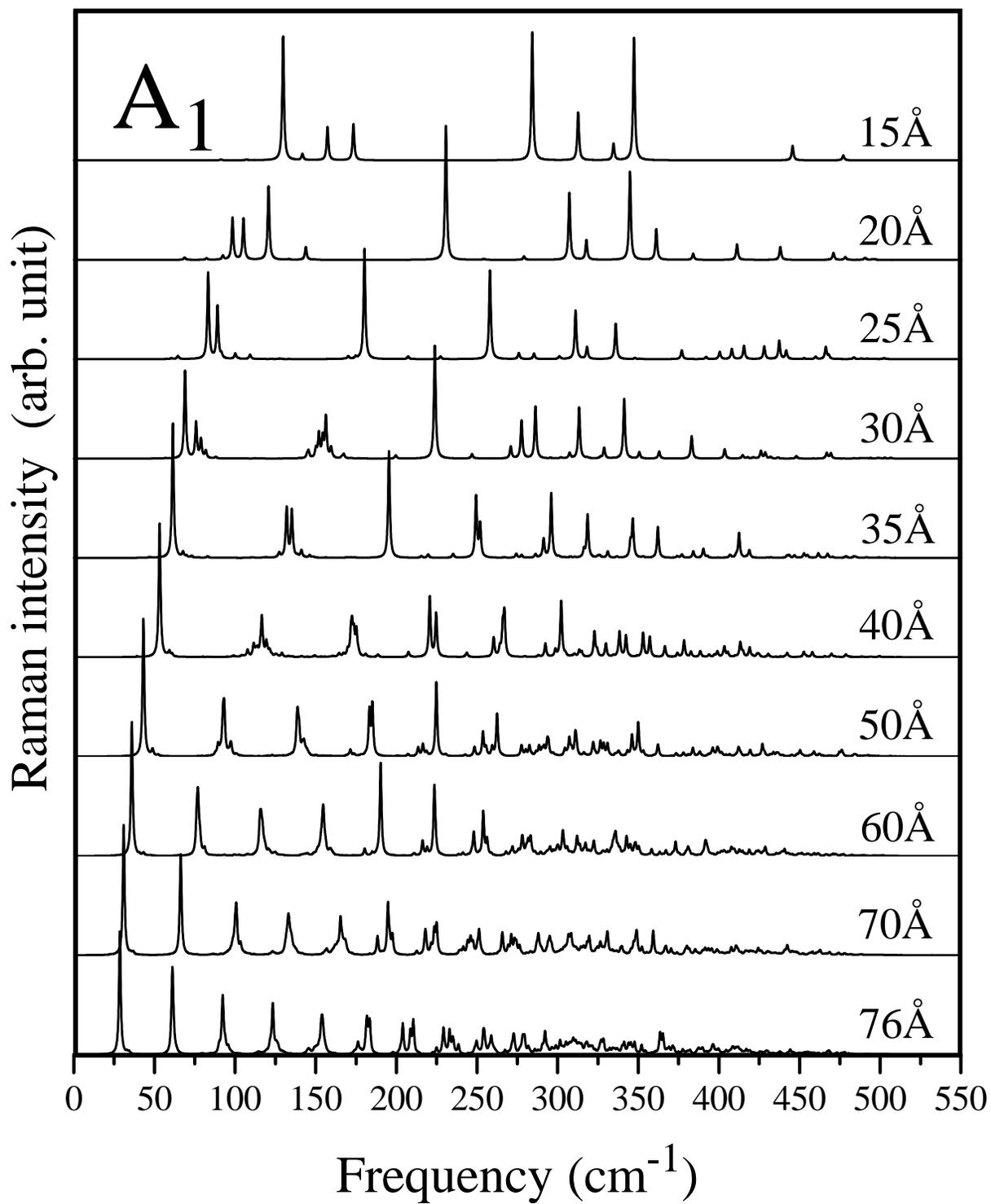
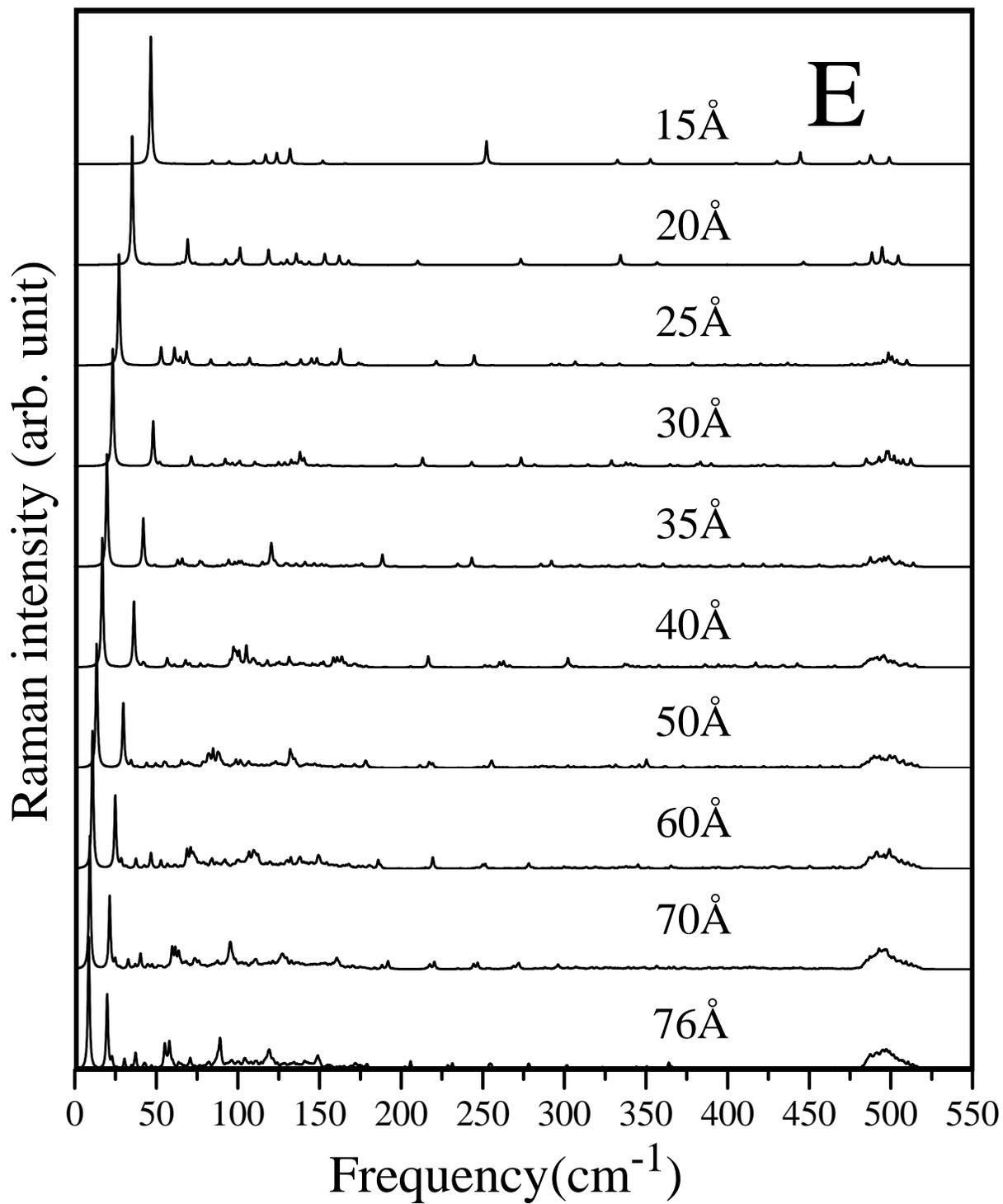
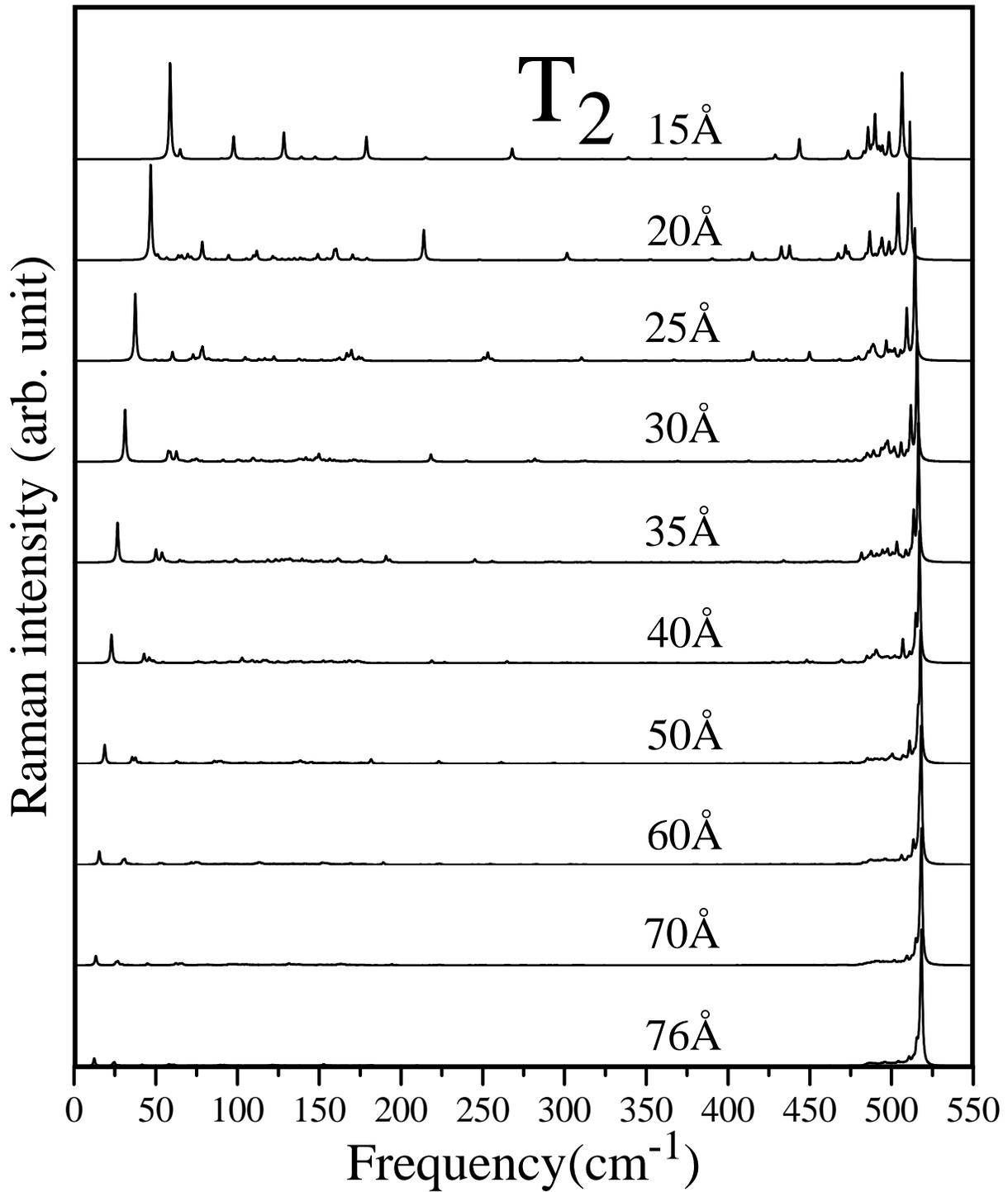
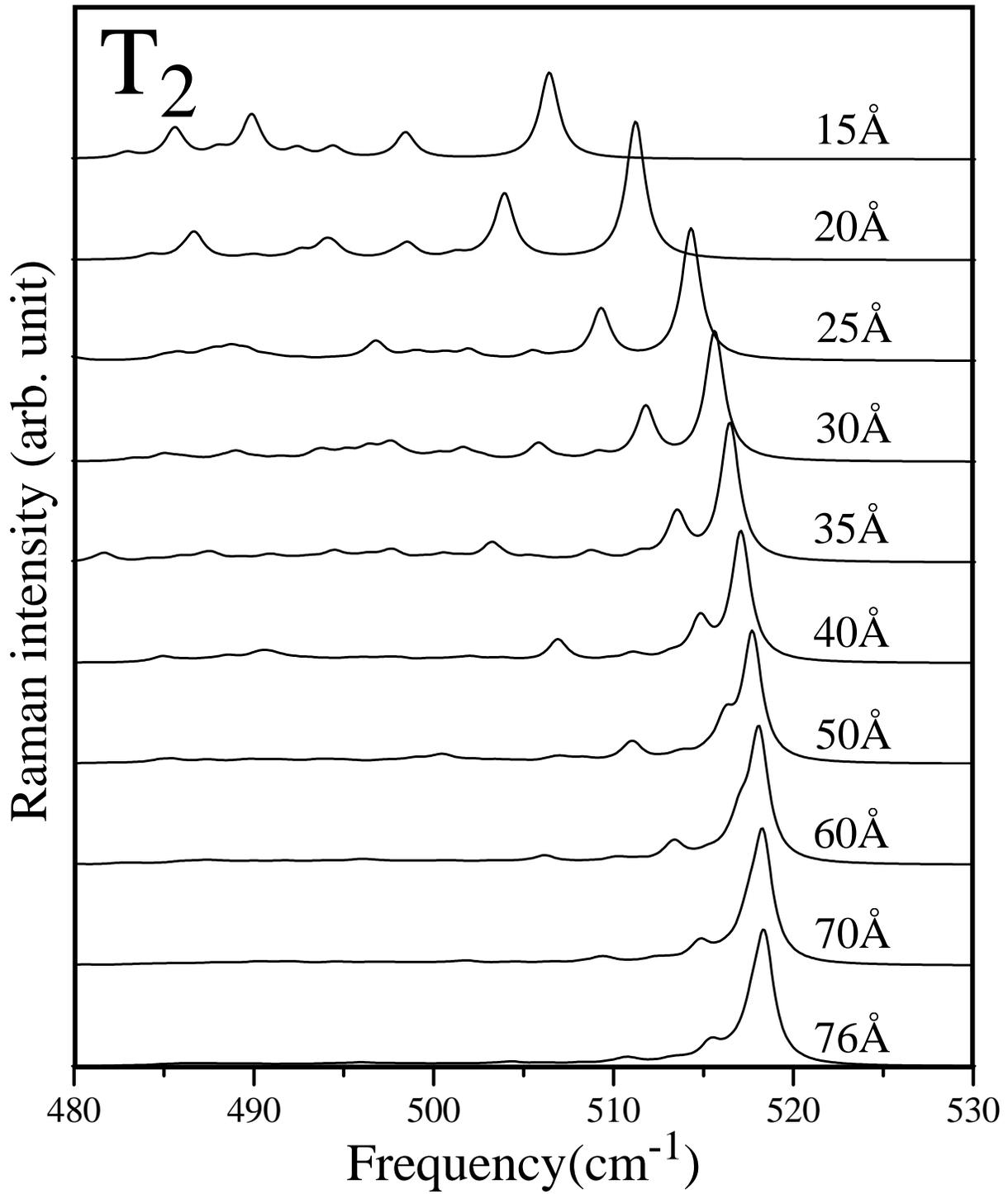
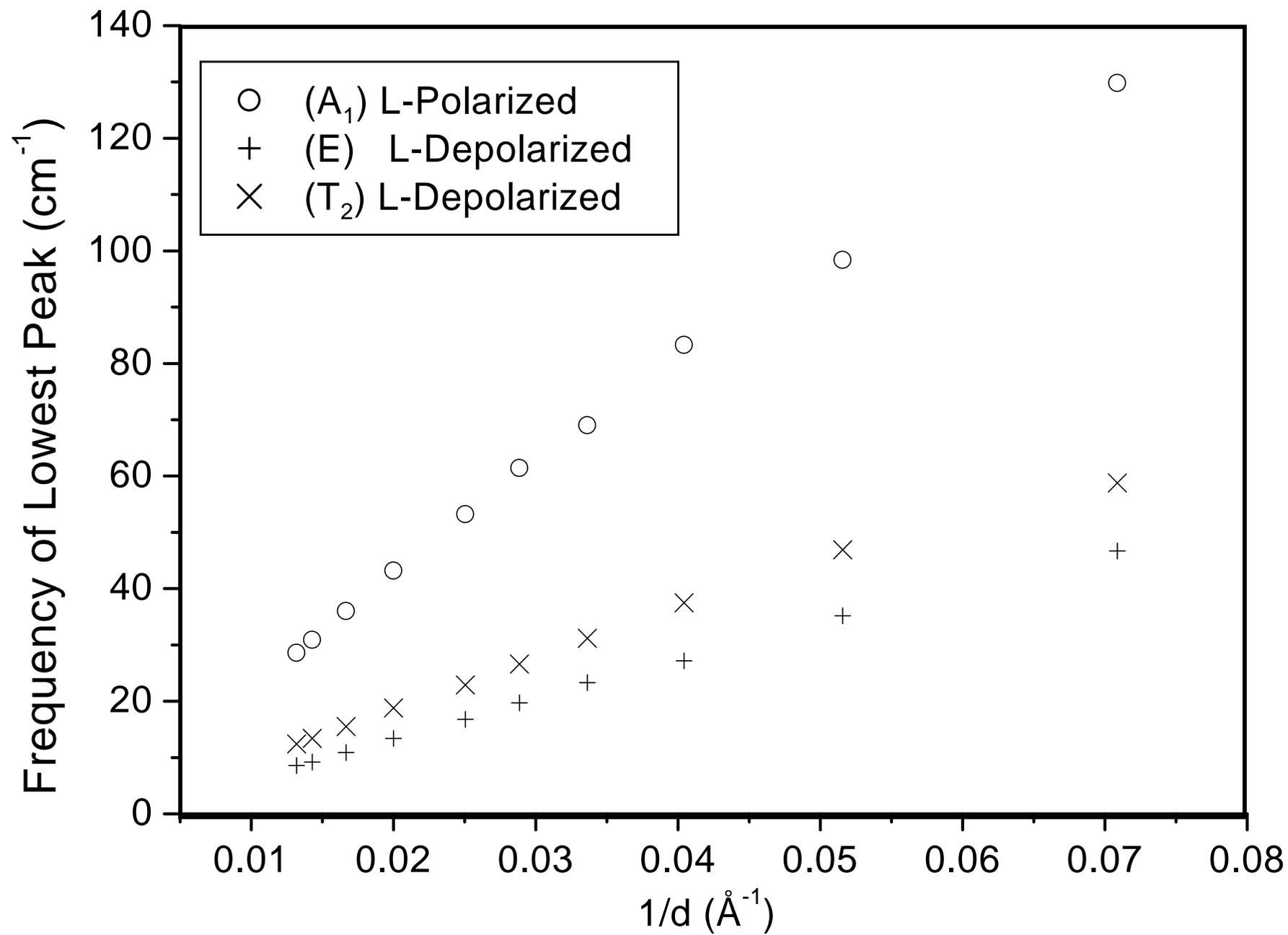
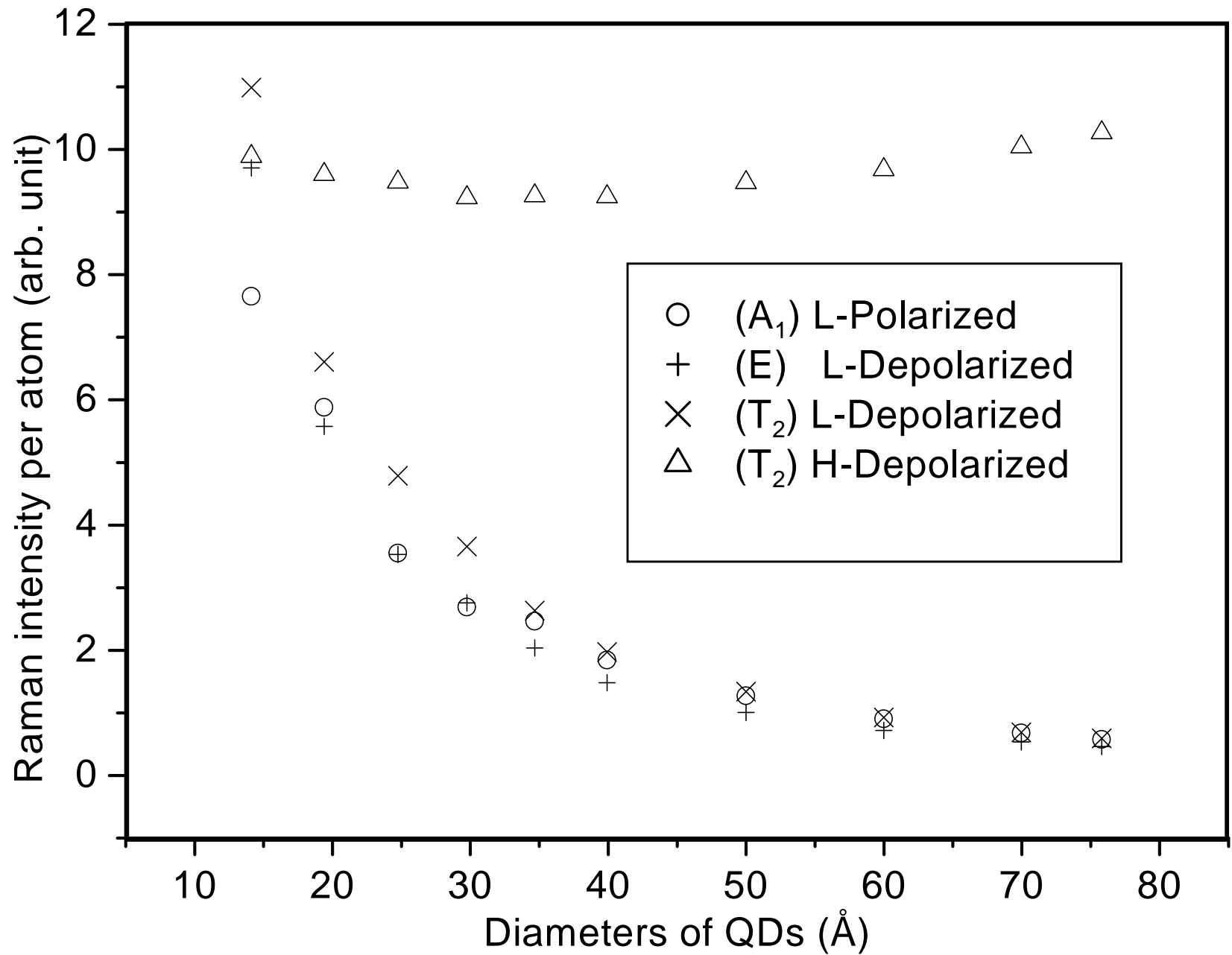
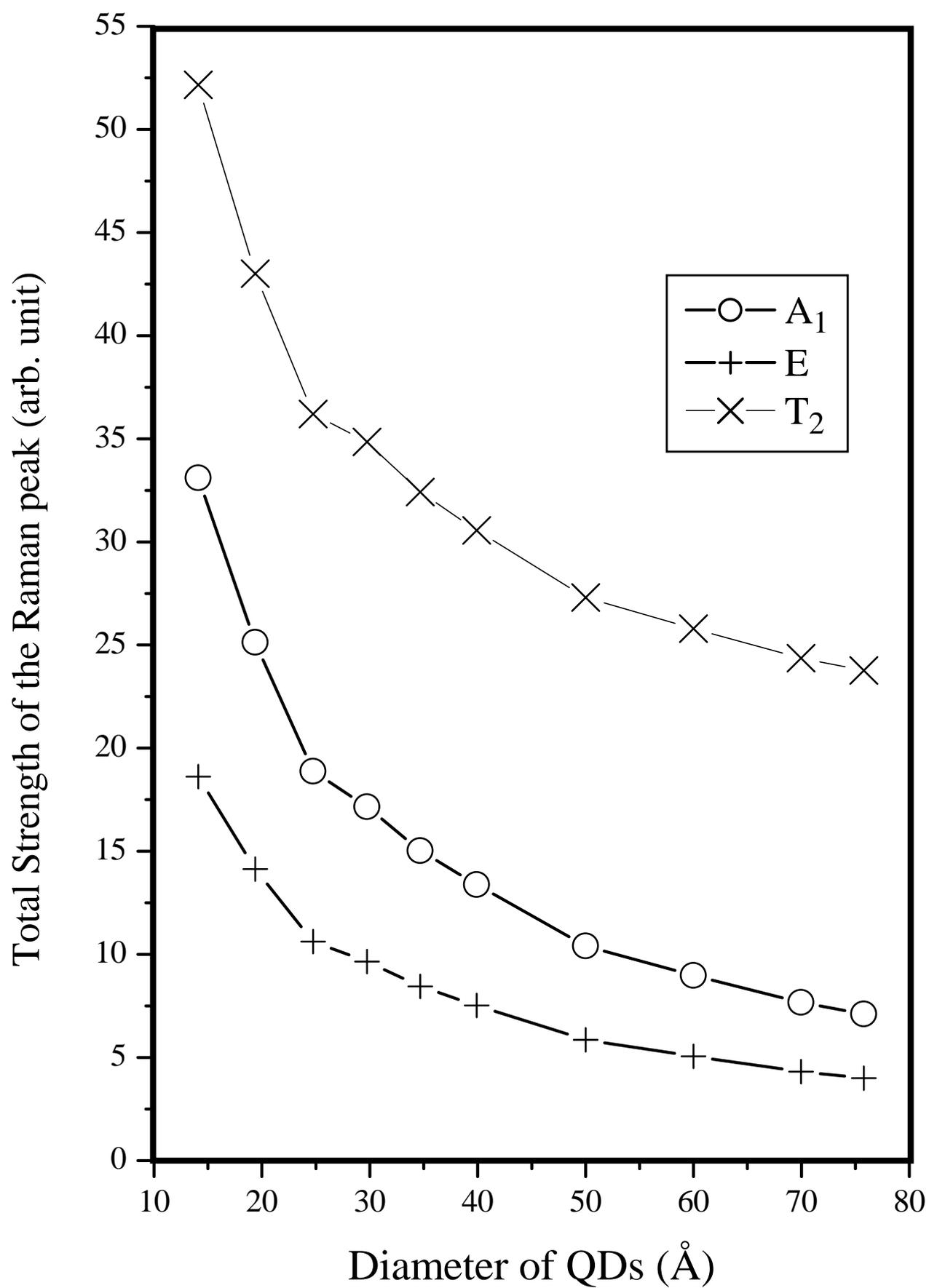
\end{document}